# Statistical Analysis for Component Shift in Pick and Place Process of Surface Mount Technology

Shun Cao[a], Irandokht Parviziomran[a], Krishnaswami Srihari[a], Seungbae Park[b], Daehan Won[a]

[a]Department of Systems Science and Industrial Engineering, The state University of New York at Binghamton, Binghamton, NY, USA

[b]Department of Mechanical Engineering, The state University of New York at Binghamton, Binghamton, NY, USA

**Abstract**

The placed electronic component can shift on the wet solder paste in pick and place (P&P) process of surface mount technology (SMT). It does not usually attract much attention, because the shift is considered to be negligibly small and the following self-alignment effect in the solder paste reflow soldering process could also make it up. However, with the decreasing size of the electronic components and the increasing demand for the low defective rate of PCB, the component shift in P&P process is becoming more and more important in quality control of SMT industries. Though a few papers are related to the component shift in P&P process, there is no earlier research using the data from the real production line. In this paper, we study two basic and important issues: the behavior of the component shift in P&P process and the contributing factors to it. Several statistical methods are used to explore the behavior component shift based on the data from a complete state-of-the-art SMT assembly line. Main effects and regression analysis are implemented to pinpoint the contributing factors. In order to investigate the issues comprehensively, six types of electronic components and multiple potential factors are considered in this work, e.g., solder paste properties (position, volume, area, height), designed position of the component, placement pressure. The results indicate that component shift cannot be ignored. Also, the position of solder paste, designed position of component and component type are the top three most important factors to study the component shifts in P&P process.

## 1. Introduction

Surface mount technology (SMT) is well-known as an essential method for electronic component assembly. The main operations in a surface mount assembly (SMA) line are stencil printing process (SPP), pick and place (P&P) and reflow soldering (for detail, see Fig. 1). A PCB stencil is aligned on the surface of the boards and solder paste is applied using a squeegee blade to ensure the pads are coated with a controlled amount of solder paste. Then, the components are mounted onto the PCB boards in their respective positions by a P&P machine. Finally, the boards are passed through a reflow oven, in which the flux in the solder paste will evaporate and the solder paste will be melt into liquid, and then form solder joints to fix the electronic components on PCB boards.

Except the three main processes mentioned above, the PCB boards in an SMA line need to be tested by a Solder Paste Inspection (SPI) machine, and one or two Automated Optical Inspection (AOI) machines to evaluate the quality of outcome for respective process. Mainly, SPI checks the quality of the solder paste after printing; Pre-AOI, which is located prior to the oven, takes charge of testing the placed components after P&P process. In terms of Post-AOI, it is laid following the oven and more accessible than Pre-AOI, which exams the condition of the assembled components after reflow soldering.

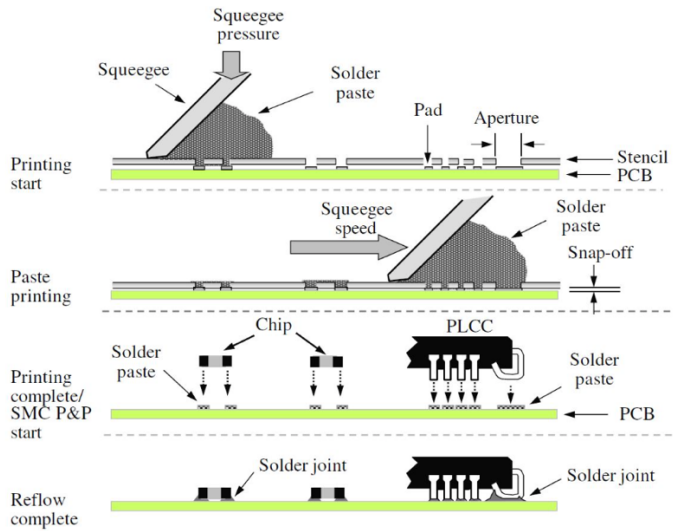

Fig. 1. Main processes of surface mount technology [1].

The P&P operations are applied on a Surface Mount Device (SMD), it is widely used in the electronics industry. When the component is placed on the wet solder paste, which could be considered as a temporary adhesive [2], it is to be held in position by the tackiness of solder paste [3]. However, the position of the placed component might be changed before entering the heating stage. The wet solder paste is a viscous non-Newtonian fluid, it will slump more or less during the P&P process. Also, if the component is placed poorly on the solder paste or the solder paste is too thin, too thick or too unbalanced etc., other forces will act on the placed component and could cause the component to shift on the wet solder paste. Besides, many other indirectly potential factors can lead to component shifts, such as machine's vibration, PCB board's oblique, conveyor's instability.

In practice, component shifts in P&P process are often underestimated because its tiny amount of the measured distance and the following reflow soldering is considered as a standard way to make up the shifts and placement errors [4]. However, with the desire of low defective rate in SMT, and the decreasing size of the electronic components, component shift in P&P process is no more ignorable, e.g., the tiny shifts may bring out significant misalignment for the small components. Note that it is challenging to detect exact value of component shift in P&P process by AOI machine in the SMA line since the difference between designed and tested positions of the component are tangled with the variations from the P&P process, inaccuracy from the inspection equipment, and hidden environmental factors (e.g., temperature, humidity). In this paper, a comprehensive experiment, which considers a variety of possible situations, even some of them rarely happens in practice, is designed for this study. We present statistical methods to analyze the behavior of component shift during the P&P process by means of the data acquired from an SMA line. And the main effects and regression analysis are implemented to pinpoint the contributing factors of the component shift.

The rest of this paper is structured as follows: lab introduction and design of experiments are briefly introduced in Sec. 2; the detail of datasets and all the factors used in this paper are described in Sec. 3; statistical analysis and results are discussed in Secs. 4 and 5; and finally, conclusions are summarized in Sec. 6.

## 2. Experimental Setup and Data Description

The experiment is carried out in the field laboratory, which contains a complete state-of-the-art SMT assembly line. The schematic diagram of the SMT line is shown in Fig. 2.

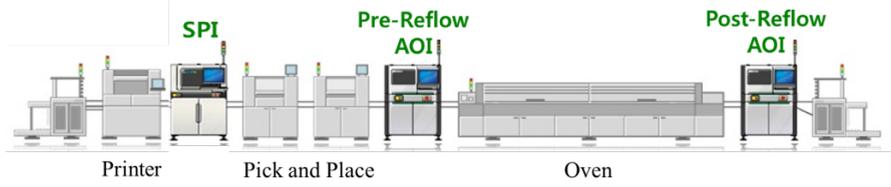

Fig. 2. Surface mount technology line in the laboratory.

Specifically, 6 types of component, 33 different settings and 20 replications of each setting are considered in this experiment, based on a full quadratic model using design of experiments. The information of components is shown in Table 1. In Table 2, we represent detailed data of the experimental settings for the capacitor 0402 as a case example.

Table 1. Components used in the experiment.

| Components | Dimensions ($\mu m$) | Quantity/board |
|---|---|---|
| Resistor Chip 01005 | 400×200 | 660 |
| Resistor Chip 0201 | 600×300 | 660 |
| Resistor Chip 0402 | 1000×500 | 660 |
| Capacitor Chip 01005 | 400×200 | 660 |
| Capacitor Chip 0201 | 600×300 | 660 |
| Capacitor Chip 0402 | 1000×500 | 660 |

Table 2. Experimental settings of Capacitor Chip 0402 for the experiment.

| Factor | Setting 1 | Setting 2 | Setting 3 | Setting 4 | Setting 5 | … | Setting 33 |
|---|---|---|---|---|---|---|---|
| Solder paste designed offset $X$ ($\mu m$) | 76.84 | 76.84 | 76.84 | 65.92 | 175.00 | … | 141.76 |
| Solder paste designed offset $Y$ ($\mu m$) | 71.12 | 71.12 | 71.12 | 129.56 | 84.00 | … | 220.23 |
| Solder paste designed Angle (°) | -6.92 | 6.92 | 6.92 | -6.92 | 6.92 | … | 6.91 |
| Average volume (%) | 80.00 | 120.00 | 120.00 | 80.00 | 120.00 | … | 120.00 |
| Difference in volume (%) | 0.00 | -40.00 | 0.00 | -40.00 | -40.00 | … | 0.00 |
| Part designed offset $X$ ($\mu m$) | 235.37 | 158.43 | 76.85 | 81.49 | 253.96 | … | 170.73 |
| Part designed offset $Y$ ($\mu m$) | 0.00 | 0.00 | 71.12 | 0.00 | 94.36 | … | 111.81 |
| Part designed Angle (°) | -6.92 | 0.00 | 0.00 | 0.00 | 0.00 | … | 0.00 |
| Placement pressure (gram-force) | 150.00 | 0.00 | 150.00 | 150.00 | 0.00 | … | 150.00 |

The datasets acquired from the SMA line are SPI dataset (size: 7920, one component corresponds to two parts of solder paste on two pads respectively) and Pre-AOI dataset (size: 3960). All the potential factors are shown in Table 3, in which the offsets are defined as the distances based on the centers: pad center (a pair of pads), solder paste center (a pair of solder paste) and component center. We illustrate the way of measuring distances in Fig. 3(a) and Fig. 3(b). The definition of relative distance is shown in Fig. 3(c). Except that, the ratio of the factor is used in the regression analysis, whereas the actual value of the factor is used in section 4 and section 5.1.

Table 3. Brief explanation of the factors used in this paper.

| Factor | Factor Name | Explanation |
|---|---|---|
| Solder paste offset $X$ ratio | $X_1$ | Solder paste offset in $X$ direction (SPI) /component length |
| Solder paste offset $Y$ ratio | $X_2$ | Solder paste offset in $Y$ direction (SPI) /component width |
| Solder paste Angle | $X_3$ | Solder paste rotation (SPI) |
| Volume ratio on left side | $X_4$ | Volume ratio of solder paste on the left side (SPI) |
| Volume ratio on right side | $X_5$ | Volume ratio of solder paste on the right side (SPI) |
| Average volume ratio | $X_6$ | Average volume ratio of a pair of solder paste |
| Difference in volume ratio | $X_7$ | Difference in volume ratio between a pair of solder paste |
| Area ratio on left side | $X_8$ | Area of solder paste on the left side (SPI)/pad area |
| Area ratio on right side | $X_9$ | Area of solder paste on the right side (SPI) /pad area |
| Average area ratio | $X_{10}$ | Average area ratio of a pair of solder paste |
| Difference in area ratio | $X_{11}$ | Difference in area ratio between a pair of solder paste |
| Height on left side | $X_{12}$ | Height of solder paste on the left side (SPI, designed heights are the same) |
| Height on right side | $X_{13}$ | Height of solder paste on the right side (SPI, designed heights are the same) |
| Part designed offset $X$ ratio | $X_{14}$ | Component designed offset in $X$ direction (DOE) /component length |
| Part designed offset $Y$ ratio | $X_{15}$ | Component designed offset in $Y$ direction (DOE) /component width |
| Part designed angle | $X_{16}$ | Component designed angle (DOE) |
| Placement pressure | $X_{17}$ | Pressure setting in the P&P machine (DOE) |
| Relative $X$ ratio | $X_{18}$ | (Part designed offset $X$ (DOE) − Solder paste offset $X$ (SPI)) /component length |
| Relative $Y$ ratio | $X_{19}$ | (Part designed offset $Y$ (DOE) − Solder paste offset $Y$ (SPI)) /component width |
| Relative Angle | $X_{20}$ | Part design Angle (DOE) − Solder paste Angle (SPI) |
| Contact area ratio on left side | $X_{21}$ | Overlap between solder paste and component (DOE) on left side/pad area |
| Contact area ratio on right side | $X_{22}$ | Overlap between solder paste and component (DOE) on right side/pad area |
| Component type | $X_{23}$ | Component type (categorical) |
| Shift $X$ | $Y_x$ | Part tested offset $X$ ratio (Pre-AOI) – Part designed offset $X$ ratio (DOE) |
| Shift $Y$ | $Y_y$ | Part tested offset $Y$ ratio (Pre-AOI) – Part designed offset $Y$ ratio (DOE) |
| Shift Angle | $Y_{ang}$ | Part tested Angle (Pre-AOI) – Part designed Angle (DOE) |

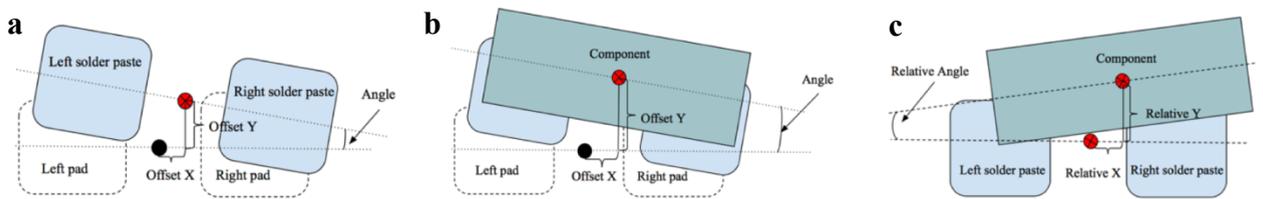

Fig. 3. (a) offsets of the solder paste; (b) offsets of the component; (c) relative distances between the solder paste and the component.

## 3. The Behavior of Component Position in Pick and Place Process

For each experimental setting, we repeat 20 placements of the same type of the components on one PCB board to verify the stability and repeatability. In Fig. 4, the black rectangle represents the designed position of the capacitor chip 0402 in setting 1, and the light blue rectangles signify the tested positions of 20 placements with the same setting.

Note that the dimension and position of the rectangles in the figure are depicted according to the actual values in the dataset. Fig. 5 shows the distribution of the shifts (or differences) between the designed and tested offsets.

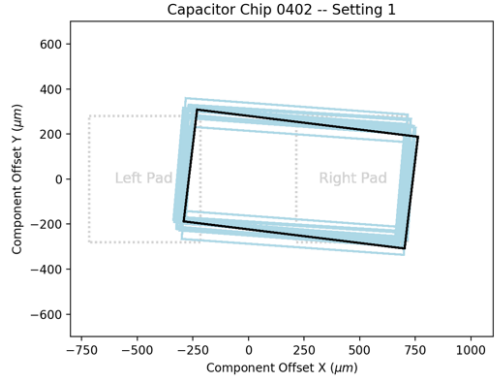

Fig. 4. Behavior of component position.

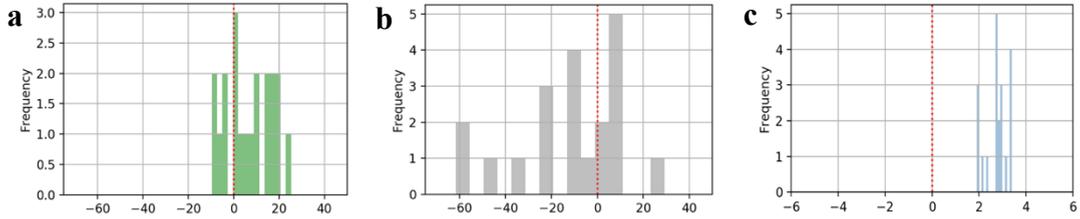

Fig. 5. (a) histogram of shift $X$ ($\mu m$); (b) histogram of shift $Y$ ($\mu m$); (c) histogram of shift Angle (°).

As shown in Fig. 4, Fig 5 and Table 4, component shifts in P&P process is still significantly considerable. A few of shift $Y$s are even greater than 10% of the component's width. Moreover, the variation of shift $Y$ is also quite large, whereas shift $X$ and Angle are relative stable because their standard deviations are smaller and vary little over different settings. However, it is noted that a certain shift $X$ and Angle are also having large distances from the expected location. For example, in Fig. 5(c), the average shift of the angles is 2.7° and substantially different from the ideal case (=0°).

Table 4. Datasets acquired from SMA line.

| Setting | Shift X ($\mu m$) | | | | Shift Y ($\mu m$) | | | | Shift Angle (°) | | | |
|---|---|---|---|---|---|---|---|---|---|---|---|---|
| | Avg. | Std. | Min. | Max. | Avg. | Std. | Min. | Max. | Avg. | Std. | Min. | Max. |
| 1 | 6.8 | 9.9 | -9.7 | 25.30 | -12.4 | 22.6 | -61.6 | 29.2 | 2.7 | 0.5 | 1.9 | 3.4 |
| 2 | -12.9 | 8.8 | -29.7 | 15.7 | 5.8 | 18.8 | -25.2 | 43.8 | -0.6 | 0.6 | -1.4 | 0.7 |
| 3 | -8.2 | 11.0 | -22.7 | 19.4 | -10.9 | 9.8 | -26.1 | 4.3 | -0.45 | 0.5 | -1.4 | 0.0 |
| 4 | -0.6 | 8.8 | -20.6 | 14.5 | -15.9 | 21.8 | -73.7 | 33.5 | -0.5 | 0.9 | -2.4 | 1.3 |
| … | … | … | … | … | … | … | … | … | … | … | … | … |
| 33 | -7.8 | 9.5 | -22.0 | 13.2 | -2.4 | 21.7 | -46.1 | 48.1 | -0.7 | 0.5 | -1.4 | 0.0 |

## 4. Contributing Factors of Component Shift in P&P Process

From the behavior of the component's position, it shows that the component shift in P&P process is worthy of attention. In this section, we focus on two issues: (i) the factors cause or are highly related to the component shifts; (ii) the relationship between these factors and the component shifts. We test the factors' effects on the component shifts firstly. Then we carry out the factor analysis by building mathematical models. Which, on one hand, can be used to decide the most contributing factors and on the other hand it can build the relationship between the factors and the component shifts. The main effects analysis and the multiple polynomial regression method are used to test the effect of each factor and to build the mathematical model respectively.

*4.1. Main effects analysis*

Main Effects plot can show how the average of response varies over the levels investigated for a certain factor. The $x$-axis in the main effects plots is marked by the levels, which are designed in the experiment. The average values of all samples studied at the respective levels are plotted on the $y$-axis. Due to space limitation, we select the main effects plots of the biggest capacitor C0402 and show it in Fig. 6. Other components show similar results to C0402. It is obvious to see that the offsets of the solder paste, the designed offsets of the component and average volume contribute more to the component shifts, which vary little with the change of the difference in volume and placement pressure.

An interesting thing could be found in Fig. 6(c) is the component's designed Angle & component shift Angle displays a nearly linear relationship. This indicates when the designed Angle of the component increases, the component shift Angle will decrease linearly.

*4.2. Regression analysis*

To verify the identification for the contribution of the factors and the responses numerically, we consider multivariate regression analysis [5]. As a preprocessing of the data collected from the lab experiment, we first remove the outliers regarding our three responses: shift $X$, shift $Y$ and shift Angle. To detect the outlier, we use Tukey's fences which is the most popular statistical technique to set the interquartile range of the acceptance region of the data Secondly, we standardize the data to reduce multicollinearity, especially for higher-order polynomial regression term. Lastly, in order to consider different sizes of components into the same scale, we use ratios of the offsets, shifts, solder paste area, and the solder paste volume, instead of actual values. All the factors and responses used in the multivariate regression analysis are shown in Table 3. In particular, two polynomial models are used in the paper; full quadratic model, and full cubic model respectively.

$R^2$, adjusted $R^2$ and predicted $R^2$ are used to measures the proportion of variability in component shift that can be explained by the factors. We represent the results of the regression analysis in Table 5. The variability of shift Angle is explained well in both the full quadratic model and the full cubic model, the $R^2$, adjusted $R^2$ and predicted $R^2$ are close to or greater than 70%. However, the variance of shift $X$ and shift $Y$ are not addressed well in the quadratic model, in which the three types of $R^2$ values are no more than 40%. While in the cubic model, the variability of shift $X$ and shift $Y$ are explained much better, but still not as good as shift Angle. Besides, the five most contributing factors of full cubic model and their respective F-values and P-values are listed in Table 6, which are ranked by the F-value.

Table 5. Regression results.

| Model | Response | $R^2$ | $R^2$ (adj) | $R^2$ (pred) |
|---|---|---|---|---|
| Full quadratic polynomial model | Shift $X$ ratio | 39.71% | 38.53% | 37.31% |
| | Shift $Y$ ratio | 39.71% | 38.72% | 37.65% |
| | Shift Angle | 70.44% | 69.99% | 69.41% |
| Full cubic polynomial model | Shift $X$ ratio | 61.44% | 59.21% | 56.47% |
| | Shift $Y$ ratio | 64.03% | 62.09% | 59.95% |
| | Shift Angle | 77.67% | 76.58% | 75.07% |

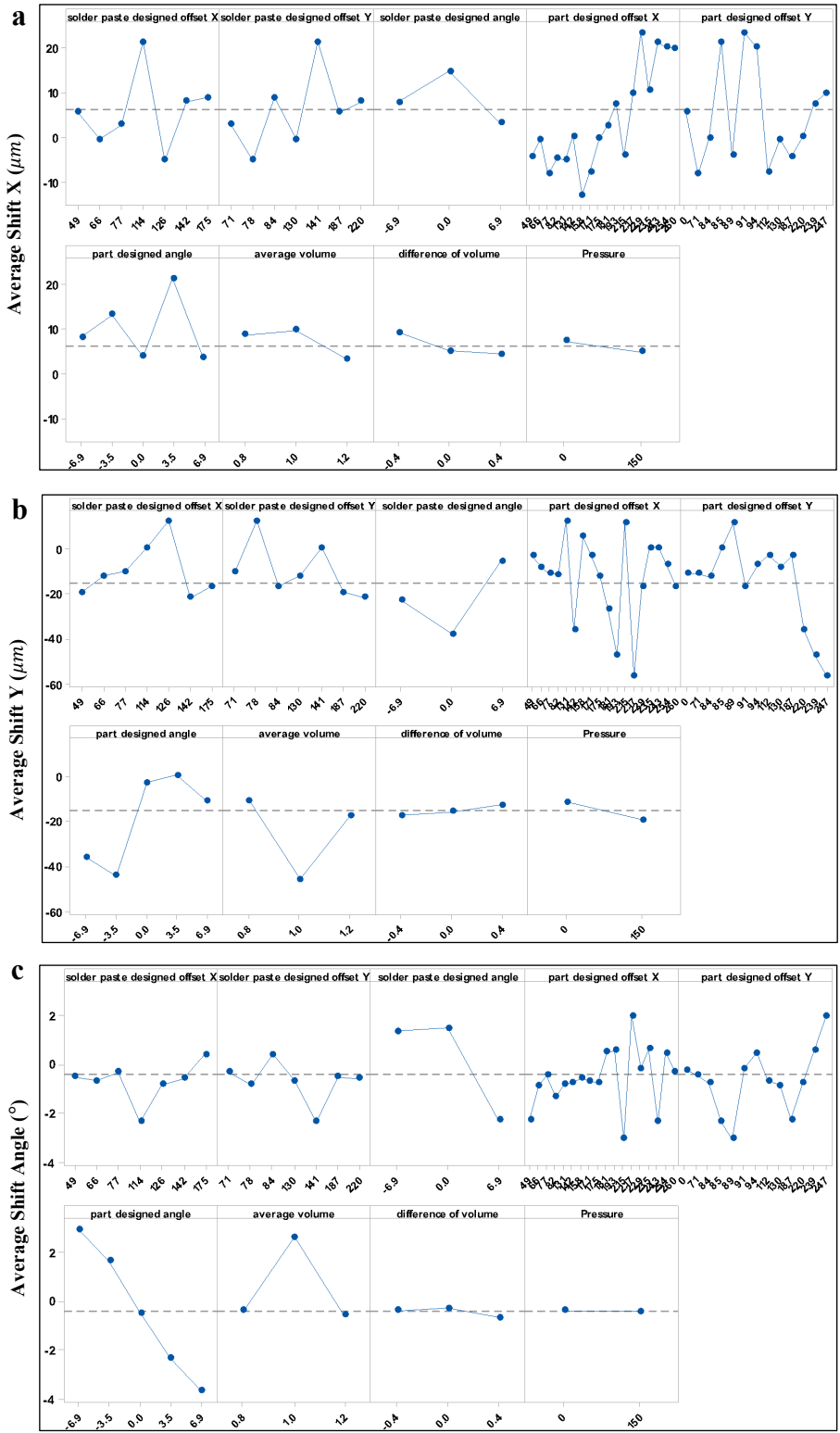

Fig. 6. (a) main effects plots of shift *X* (C0402); (b) main effects plots of shift *Y* (C0402); (c) main effects plots of shift Angle (C0402).

Table 6. Five most contributing factors in full cubic regression results.

| Shift X | | | Shift Y | | | Shift Angle | | |
|---|---|---|---|---|---|---|---|---|
| Factors | F | P | Factors | F | P | Factors | F | P |
| Relative X ratio × Relative Angle | 206 | 0.00 | Part designed Angle × Part designed Angle × Part designed Angle | 150 | 0.00 | Part designed Angle | 1014 | 0.00 |
| Contact area ratio on right side × Component type | 164 | 0.00 | Part designed Angle | 139 | 0.00 | Part designed offset Y ratio × Part designed Angle × Component type | 164 | 0.00 |
| Part designed offset Y ratio × Part designed offset Y ratio × Component type | 134 | 0.00 | Part designed offset Y ratio × Part designed Angle × Component type | 135 | 0.00 | Part designed offset Y ratio × Volume ratio on right side × Component type | 110 | 0.00 |
| Part designed offset X ratio × Contact area ratio on right side × Component type | 128 | 0.00 | Part designed Angle × Average area ratio × Difference in area ratio | 134 | 0.00 | Contact area ratio on right side × Relative X ratio × Relative Y ratio | 89 | 0.00 |
| Component type | 122 | 0.00 | Average area ratio × Placement pressure × Component type | 128 | 0.00 | Contact area ratio on right side × Difference in volume ratio × Difference in area ratio | 83 | 0.00 |

## 5. Conclusions

In this paper, we have studied the component shift: $x$, $y$, and rotation (e.g., angular variation) in P&P process of SMT. According to the data acquired from the lab experiment, it can be shown that the component shift is not as small as to be neglected since a certain amount of the shifts are distributed far from the expected position. Particularly, a few of shift $Y$s are more than 10% of component's width. The results of main effects and regression analysis indicate that the position of solder paste, the designed position of the component, relative distances and average volume of solder paste have the significant influence on the component shift occurred in P&P process. Also, the component type is another important factor which can be seen in Table 6. Besides, shift angle is nearly negative proportional with the designed angle of the component. The preliminary results are valuable for the study of component shift in P&P process. Although we could determine the contributing factors, it is still challenging to state these factors will definitely cause the component to shift. To establish a clear determination of the shifts, we will extend the scope of the involved factors to machinery and inspectional accuracy. Thus, it will encompass more factors to better describe the component shift in P&P process and help us to reduce the unexpected shift which may cause products' defects.

## Acknowledgments

We would like to thank the editors and the anonymous reviewers for your comments.## References

[1] Tsai, Tsung-Nan. "Modeling and optimization of stencil printing operations: A comparison study." Computers & Industrial Engineering 54.3 (2008): 374-389.
[2] Hwang, Jennie S. Solder paste in electronics packaging: technology and applications in surface mount, hybrid circuits, and component assembly. Springer Science & Business Media, 2012
[3] Liukkonen, Timo, Pekka Nummenpää, and Aulis Tuominen. "The effect of lead-free solder paste on component placement accuracy and self-alignment during reflow." Soldering & surface mount technology 16.1 (2004): 44-47.
[4] Lee, Ning-Cheng. Reflow Soldering Processes. Elsevier, 2002
[5] Darlington, Richard B., and Andrew F. Hayes. Regression analysis and linear models: Concepts, applications, and implementation. Guilford Publications, 2016.